\documentstyle[12pt]{article}

\begin{document}
\thispagestyle{empty}

\begin{center}
               RUSSIAN GRAVITATIONAL SOCIETY\\
               INSTITUTE OF METROLOGICAL SERVICE \\
               CENTER OF GRAVITATION AND FUNDAMENTAL METROLOGY\\

\end{center}
\vskip 4ex
\begin{flushright}
                                         RGS-VNIIMS-004/98
                                         \\ gr-qc/9810019

\end{flushright}
\vskip 15mm

\begin{center}
{\large\bf Superluminal propagation of light in gravitational field and
non-causal signals: some comments\footnote{ This work was supported by the
Russian Ministry of Science and the Russian Foundation of Basic Research
(grant N 98-02-16414).}
}

\vskip 5mm
{\bf
M.Yu. Konstantinov }\\
\vskip 5mm
     {\em VNIIMS, 3-1 M. Uljanovoj str., Moscow, 117313, Russia}\\
     e-mail: konst@rgs.phys.msu.su \\
\end{center}
\vskip 10mm

\begin{abstract}
We examine the recent statement by A. Dolgov and I. Novikov that
superluminal propagation of light, which is induced by the quantum
corrections to the photon propagation in gravitational field, permits
non-causal signal propagation. For this purpose we examine the possibility
of the existence of non-causal signals in the model case, when
characteristic cone of the signal is outside the usual light cone in
Minkowski space-time, as it take place in the case of the photon propagation
in gravitation field with quantum corrections taking into account. In such
model the signal propagates along invariant intervals. It is shown that
there are no non-causal signals in the considered model. The recent
statement of A. Dolgov and I. Novikov was obtained by using additional
supposition that velocity of the signal with respect to the emitter is
independent from its motion. Such supposition does not valid in the case of
the photon propagation in gravitational field.
\end{abstract}

\vskip 10mm

%PACS numbers:   \\

\vskip 30mm

\centerline{\bf Moscow 1998}
\pagebreak

%%%%%%%%%%%%%%%%%%%%%%%%%%%%%%%%%%%%%%%%%%%%%%%%%%%%%%%%%%%%%%%%%%

\section{Introduction}

The recent studies of photon propagation in gravitational field show that
quantum corrections modify the characteristic cone of electromagnetic field
equations in such a way that in some cases it may lay outside the light cone~%
\cite{dh,ds1,khripl,shor,ds2,moha}. This effect follows from the photon
effective action in gravitational field in one loop approximation, which has
the following form~\cite{ds1,shor,ds2}

\begin{equation}
\label{s}S=\int d^4x\sqrt{-g}\left( -{\frac 14}F_{\mu \nu }F^{\mu \nu
}+\frac \alpha {m_e^2}\left( aRF_{\mu \nu }F^{\mu \nu }+bR_{\mu \nu }F^{\mu
\lambda }F_\lambda ^\nu +cR_{\alpha \beta \mu \nu }F^{\mu \nu }F^{\alpha
\beta }\right) \right)
\end{equation}
Here $F_{\mu \nu }=\partial _\mu A_\nu -\partial _\nu A_\mu $ is the
electromagnetic field tensor, $\alpha =1/137$ is the fine structure
constant, $m_e$ is the mass of electron and $a$, $b$ and $c$ are
coefficients, whose exact values are calculated in~\cite{dh,ds1,shor,ds2}.
For our purpose it is essential that these coefficients are nonzero. The
terms, which are proportional to the Ricci tensor $R_{\mu \nu }$ or to the
curvature scalar $R$ vanish in vacuum. It is evident that the action~(\ref{s}%
) leads to the field equations whose characteristic cone does not coincide
with the normal light cone in general case. It means that velocity of the
front of the signal would not coincide with the normal light velocity $c$.
Geometrically it means that instead of the normal isotropic interval in
normal metric
\begin{equation}
\label{light}ds^2=g_{\mu \nu }dx^\mu dx^\nu ,
\end{equation}
photon propagate along interval whose line element $ds_{ph}^2$ may be
positive, negative or zero depending from direction of propagation,
polarization etc. The case $ds_{ph}^2<0$ corresponds to faster than light
photon propagation\footnote{%
It is supposed that space-time metric has signature $(+,-,-,-)$.} (examples
see in~\cite{dh,ds1,khripl,shor,ds2,moha}). In some cases such nonisotropic
photon propagation in normal metric may be described as propagation along
isotropic direction in modified metric
\begin{equation}
\label{modifint}d\widetilde{s}^2=\widetilde{g}_{\mu \nu }dx^\mu dx^\nu
\end{equation}
where $\widetilde{g}_{\mu \nu }$ is effective metric tensor.

The most important that photon propagation is determined by general
covariant equations and hence it propagate along invariant curve, which may
be space-like in normal metric~(\ref{light}) and does not depend on the
emitter motion. The possibility to describe photon propagation by means of
isotropic curves in modified metric~(\ref{modifint}) is not essential.

In the standard Special Relativity it is usually expected (see, for instance~%
\cite{dn,hawking}), that ''time travel and faster than light space travel
are closely connected'' and ''if you can do one you can do another''. In few
words such connection is explained as follows~\cite{hawking}: ''You just
have to travel from $A$ to $B$ faster than light would normally take. You
then travel back, again faster than light, but in a different Lorentz frame.
You can arrive back before you left'' (in more details this discussed in~%
\cite{dn}). Therefore, the faster than light photon propagation ''may imply
serious problems for the theory''~\cite{dn}.

In this note we examine the possibility of the appearing of non-causal
signals, which declared~\cite{dn}, in the model case than the superluminal
propagation of signal occurs in flat space-time. As in the case of the
superluminal photon propagation in gravitational field, it will be assumed
that the superluminal signal propagate along isotropic line in some
effective metric, which is not coincide with normal Lorentzian metric. It
will be shown that the non-causal signal propagation in this case is
impossible.

\section{Superluminal signals in Special Relativity}

Consider the superlight signal propagation in Special Relativity. Namely,
consider the usual Minkowski $\left( t,x\right) $-plane with interval
\begin{equation}
\label{minkint}ds^2=dt^2-dx^2
\end{equation}
and suppose that it is possible to send superluminal signals, whose interval
has the form
\begin{equation}
\label{tachyon}d\widetilde{s}^2=u^2dt^2-dx^2=0
\end{equation}
where $u>1$. Such supposition may be considered as a simplest model for
superluminal photon propagation due to the quantum corrections.

\begin{figure}
\begin{center}
\begin{picture}(70,70)
\put(33,0){\vector(0,1){65}}
\put(0,33){\vector(1,0){65}}
\put(65,34){$x$}
\put(34,65){$t$}
\put(65,65){$B'$}
\put(65,3){$A$}
\put(3,65){$A'$}
\put(5,3){$B$}
\put(33,33){\line(1,1){30}}
\put(33,33){\vector(1,1){15}}
\put(33,33){\line(1,-1){30}}
\put(33,33){\line(-1,-1){30}}
\put(33,33){\line(-1,1){30}}
\put(33,33){\vector(-1,1){15}}
\put(34,37){$p_0$}
\put(34,53){$p_2$}
\put(52,41){$p_1$}
\thicklines
\put(33,33){\line(2,1){30}}
\put(15,24){\vector(2,1){10}}
\put(33,33){\vector(2,1){10}}
\put(33,33){\line(2,-1){30}}
\put(33,33){\line(-2,1){30}}
\put(51,24){\vector(-2,1){10}}
\put(33,33){\vector(-2,1){10}}
\put(33,33){\line(-2,-1){30}}
\put(33,33){\circle*{1.5}}
\put(51,42){\circle*{1}}
\put(51,42){\vector(-2,1){18}}
\put(33,51){\circle*{1}}
\end{picture}
\caption{Suprlight photon propagation in Minkowski space-time}\label{rest}
\end{center}
\end{figure}
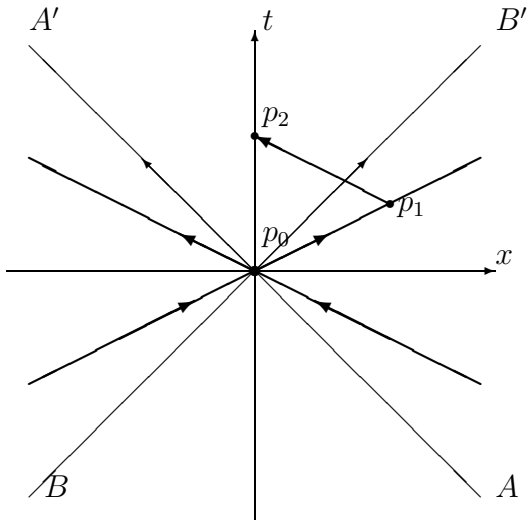

Let such superluminal signal is emitted at the point $x_0=0$ at the moment $%
t_0=0$ (the event $p_0$), then it is detected at the point $x_1$ at the
moment $t_1$ and immediately re-emitted in reverse direction (the event $p_1$%
) and finally is detected at the initial point $x_0$ at some moment $t_2$
(the event $p_2$) as it is shown at figure~\ref{rest},
where the thin lines $AA^{\prime }$ and $BB^{\prime }$ represent the normal
light cone for the Minkowski metric~(\ref{minkint}), the thick lines
represent the tachyonic cone (isotropic cone in modified metric~(\ref
{tachyon})), the arrows define the directions of the superluminal signal
propagation and $u=2$. It is clear, that intervals $p_0p_1$, $p_1p_2$ and $%
p_0p_2$ are invariants with respect to arbitrary coordinate transformations.
In particular, let us make Lorentz transformation
\begin{equation}
\label{lorentz}x^{\prime }=\beta (x-Vt),\qquad t^{\prime }=\beta (t-Vx)
\end{equation}
where $\beta =1/\sqrt{1-V^2}$. The interval~(\ref{minkint}) is invariant
under this transformations, while the interval~(\ref{tachyon}) takes the
form
\begin{equation}
\label{modtachyon}d\widetilde{s}^2=\beta ^2(adt^{\prime 2}-2bdt^{\prime
}dx^{\prime }-cdx^{\prime 2})
\end{equation}
where $a=(u^2-V^2)$, $b=(1-u^2)V$ and $c=(1-u^2V^2)$. In this moving
reference frame superluminal signal propagates with velocities
\begin{equation}
\label{velpos}u_{+}^{\prime }=\frac{u-V}{1-uV}
\end{equation}
in positive direction of the $x^{\prime }$ - axis (the possibility $%
u_{+}^{\prime }<0$ means that $t_1^{\prime }<t_0^{\prime }$) and
\begin{equation}
\label{velneg}u_{-}^{\prime }=-\frac{u+V}{1+uV}
\end{equation}
in negative direction of $x^{\prime }$ - axis. Thus, in the moving reference
frame the velocity of the signal propagation not only changes its value, but
become asymmetric with respect to space reflection.

It is clear, that equations~(\ref{velpos}) and~(\ref{velneg}) are the direct
consequence of the velocity-addition formula, which is well known from
special relativity. So, our description of superluminal signal propagation
by means of isotropic curves in modified metric~(\ref{tachyon}) is
equivalent to the description of superluminal signal propagation by means of
invariant space-like curves in normal metric~(\ref{minkint}).

\begin{figure}
\begin{center}
\begin{picture}(70,70)
\put(33,0){\vector(0,1){65}}
\put(65,34){$x'$}
\put(34,65){$t'$}
\put(16,65){$t$}
\put(65,16){$x$}
\put(65,65){$B'$}
\put(65,3){$A$}
\put(3,65){$A'$}
\put(5,3){$B$}
\put(33,33){\line(1,1){30}}
\put(33,33){\vector(2,-1){30}}
\put(33,33){\line(-2,1){30}}
\put(33,33){\vector(-1,2){16}}
\put(33,33){\line(1,-2){16}}
\put(33,33){\vector(1,1){15}}
\put(33,33){\line(1,-1){30}}
\put(33,33){\line(-1,-1){30}}
\put(33,33){\line(-1,1){30}}
\put(33,33){\vector(-1,1){15}}
\put(34,35){$p_0$}
\put(24,53){$p_2$}
\put(52,34){$p_1$}
\thicklines
\put(0,33){\vector(1,0){65}}
\put(33,33){\vector(1,0){10}}
\put(15,33){\vector(1,0){10}}
\put(33,33){\line(3,-2){30}}
\put(33,33){\vector(-3,2){10}}
\put(33,33){\line(-3,2){30}}
\put(33,33){\circle*{1.2}}
\put(51,33){\circle*{1}}
\put(51,33){\vector(-3,2){27}}
\put(51,21){\vector(-3,2){12}}
\put(24,51){\circle*{1}}
\end{picture}
\caption{Suprlight photon propagation in Minkowski
space-time in moving reference frame}\label{moving}
\end{center}
\end{figure}
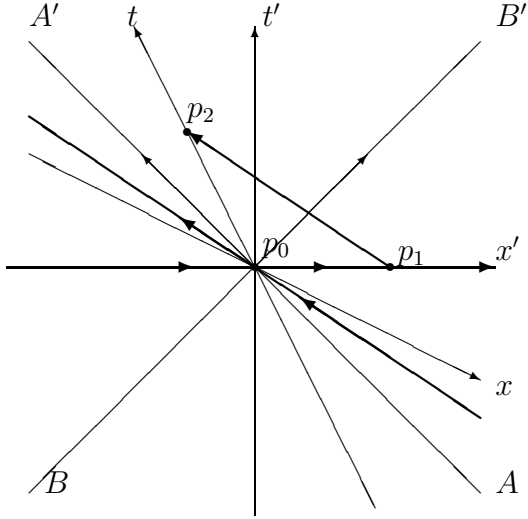

In particular, for $V=1/u$ equations~(\ref{velpos}),~(\ref{velneg}) give $%
u_{+}^{\prime }=\infty $ and $u_{-}^{\prime }=-(1+u^2)/2u$, the events $p_0$
and $p_1$ become simultaneous, i.e. $t_1^{\prime }=t_0^{\prime }$ but in
both reference systems $\left( x,t\right) $ and $\left( x^{\prime
},t^{\prime }\right) $ signal propagates from $p_0$ to $p_1$ and from $p_1$
to $p_2$ but not reverse (see figure~\ref{moving}).
In particular, if reference system $\left( x^{\prime \prime },t^{\prime
\prime }\right) $ moves with respect to the system $\left( x,t\right) $ with
velocity $V=-1/u$ than in the reference system $\left( x^{\prime \prime
},t^{\prime \prime }\right) $  the simultaneous events are $p_1$ and $p_2$
and the signal propagates from  $p_0$ to $p_1$ and from $p_1$ to $p_2$ but
not reverse. It is easy to see that there is no reference frame where the
signal propagates from $p_1$ to $p_0$ or from $p_2$ to $p_1$. So, the
appearance of non-causal signals in the considered model is impossible.

The generalization of the considered model to the superluminal photon
propagation in gravitational field, is straightforward and will not
considered here.

\section{Conclusion}

Thus it is shown that if the superluminal signal propagate along invariant
space-like interval (or along isotropic interval in modified metric), than
the non-causal signal propagation is impossible. So, the superluminal
propagation of light in gravitation field, which was studied by several
authors~\cite{dh,ds1,khripl,shor,ds2,moha}, does not lead to the appearance
of non-causal signals.

The ''standard'' conclusions (see, for instance,~\cite{dn,hawking}) about
connection between faster than light space travel and time travel, are made
in supposition that the superluminal signal has the same velocity $u$
independent from the source motion (in particular, in~\cite{dn} we find:
''Let us suppose for simplicity that the picture is symmetric so that the
second source/detector emits a tachyon with the same velocity $u$ with
respect to itself ...''). It is easy to see that this supposition, which is
natural in Newtonian mechanics, can not be applied to the generally
relativistic case of superluminal photon propagation, which induced by
quantum corrections in gravitational field. In particular, as it was shown
in the simplest model of superluminal signal propagation in Minkowski
space-time, if the equations of the signal propagation are generally
covariant, as in the case of superluminal photon propagation in
gravitational field, than the space velocity of the signal satisfies to the
well known velocity-addition formula. As a result, in the moving reference
system velocity of signal changes by such a way that non-causal signals are
impossible.

Some remarks about the above model of superluminal photon propagation are
necessary. First, this model demonstrates the principle possibility of
coexistence of several non-equivalent metric (more exact causal) structures
in the same space-time. In another context such possibility was supposed in~%
\cite{konst85}. Some cosmological implications of such possibility were
discussed in~\cite{gons-mest} and the possible geometric description of such
situation and some its implications, in particular, the Lorentz invariance
of ''standard'' fields and appearance of ''dark'' matter, were considered in~%
\cite{konst,konst-mg8}. Second, the conclusions about superluminal photon
propagation in gravitational field~\cite{dh,ds1,khripl,shor,ds2,moha} was
obtained in the test field approximation when the gravitational field is
given and independent from the electromagnetic one. It is not strange that
characteristic cones of electromagnetic and gravitational field equations
are not coincide. If the self-consistent system of gravitation and
electromagnetic fields would be considered with quantum correction to the
photon propagation taking into account than both gravitational and
electromagnetic fields would have the same characteristic cone and by this
reason the electromagnetic signal will propagate with the same velocity as
gravitational signal. Nevertheless, it is not necessary that in such
self-consistent mode the gravitational and electromagnetic signals would
propagate along isotropic cone in the metric which solves modified
Einstein-Maxwel equations. The detailed investigation of this problem
requires rather cumbersome calculations and separate consideration.

\end{document}